\documentclass[10pt,conference]{IEEEtran}

\usepackage[caption=false]{subfig}
\usepackage{color}
\usepackage{algorithmic}
\usepackage[ruled,vlined,longend,linesnumbered]{algorithm2e}
\usepackage{graphicx}
\usepackage{epstopdf}
\usepackage{tabularx,ragged2e}
\usepackage{pdfpages}
\usepackage{enumitem}
\usepackage{multirow}
\newcolumntype{L}{>{\RaggedRight\arraybackslash}X}
\usepackage{wrapfig}
\usepackage[numbers,sort&compress,square]{natbib}

\usepackage{pifont}
\PassOptionsToPackage{hyphens}{url}

\AtBeginDocument{\renewcommand{\ref}[1]{\autoref{#1}}}
\usepackage[english]{babel}
\usepackage{hyperref}
\addto\extrasenglish{%
}

\title{Swiss Cheese Model for AI Safety: A Taxonomy and Reference Architecture for Multi-Layered Guardrails of Foundation Model Based Agents}

\author{
    Md Shamsujjoha\textsuperscript{*}, Qinghua Lu, Dehai Zhao, Liming Zhu \\
    Data61, CSIRO, Australia \\
    Email: \{md.shamsujjoha, qinghua.lu, dehai.zhao, liming.zhu\}@data61.csiro.au
}

\begin{document}
\maketitle

\begingroup
\renewcommand\thefootnote{\textsuperscript{*}}
\footnotetext{\ding{41} \href{mailto:dishacse@yahoo.com}{Corresponding Author: Md. Shamsujjoha}}
\endgroup

\begin{abstract}
Foundation Model (FM)-based agents are revolutionizing application development across various domains. However, their rapidly growing capabilities and autonomy have raised significant concerns about AI safety. Researchers are exploring better ways to design guardrails to ensure that the runtime behavior of FM-based agents remains within specific boundaries. Nevertheless, designing effective runtime guardrails is challenging due to the agents' autonomous and non-deterministic behavior. The involvement of multiple pipeline stages and agent artifacts, such as goals, plans, tools, at runtime further complicates these issues. Addressing these challenges at runtime requires multi-layered guardrails that operate effectively at various levels of the agent architecture.
Therefore, in this paper, based on the results of a systematic literature review, we present a comprehensive taxonomy of runtime guardrails for FM-based agents to identify the key quality attributes for guardrails and design dimensions. Inspired by the Swiss Cheese Model, we also propose a reference architecture for designing multi-layered runtime guardrails for FM-based agents, which includes three dimensions: quality attributes, pipelines, and artifacts. The proposed taxonomy and reference architecture provide concrete and robust guidance for researchers and practitioners to build AI-safety-by-design from a software architecture perspective.

\end{abstract}


\section{Introduction}
\label{Introduction}
A \textbf{F}oundation \textbf{M}odel (FM) is a large-scale machine learning model pre-trained on massive amounts of data using self-supervision at scale. These models are highly versatile and can adapt to a wide range of downstream tasks~\cite{29-ed-foundation-model-opportunity-risk2021}. 
The term `foundation' reflects their role as the fundamental base upon which many specialized models/systems are built. However, it is important to recognize that FM-based systems exhibit inherent limitations, particularly when handling complex tasks. Users are often required to provide detailed instructions, which can lead to inefficiencies and is prone to error.

An FM-based agent is an autonomous system that is capable of perceiving context, reasoning, planning, and executing workflows by interacting with FMs, external tools, knowledge bases, and other agents to achieve human goals~\cite{qinghua-ref-architecture-2024}. 
There has been extensive interest in FM-based agent development recently due to their huge potential to enhance productivity across various domains. However, their autonomous and non-deterministic behavior introduce substantial concerns regarding AI safety~\cite{RAI-book-qinghua, bass25engineering}, such as generating harmful or offensive content, producing dangerous or unintended outcomes, spreading disinformation and misinformation, etc~\cite{boming-ai-safety-taxonomy}. 

To address these challenges, effective runtime guardrails are key to ensure that agents behave in a safe and responsible manner~\cite{RAI-book-qinghua}. In this context, guardrails are mechanisms integrated into an agent's architecture to safeguard its behavior during runtime, preventing undesirable or unsafe behaviors~\cite{bass25engineering}. 
There have been some initial efforts on runtime  guardrails such as input filtering~\cite{9-guardrails-for-chatboats,29-ed-foundation-model-opportunity-risk2021}, output modification~\cite{28-ed-decoding-trust-gpt,23-ed-safety-assessing-llm}, adaptive fail-safes~\cite{5-frontier-AI-risk,12-ed-jailbroken}, real-time monitoring and detection~\cite{AI-Guardrials,3-ed-build-guardrails-llm, 8-nemo-guardrails,31-ed-kang2020model}, and continuous output validation~\cite{15-ed-auditing-llm,19-ed-mitigate-security-llm,FM-survey-2023-bert}. 

However, the existing guardrail approaches primarily address functional correctness, often overlooking quality attributes of FM-based agents, such as customizability and interpretability. Most importantly, these approaches mainly focus on individual single-layered guardrails that are narrowly applied to specific agent artifacts, such as prompts or FM outputs, which are insufficient to manage the inherent autonomy and non-deterministic nature of FM-agents. If any single guardrail fails, the associated risks may bypass it, potentially impacting the final results of the FM-based agent.  

Therefore, in this paper, we first present a comprehensive taxonomy to categorize runtime guardrails from a software architecture perspective, based on the results of a systematic literature review. The taxonomy comprises two primary categories: quality attributes and design options. Inspired by Swiss Cheese Model~\cite{swiss-cheesem-risk-management-review}, we also propose novel reference architecture for designing multi-layered guardrails of FM-based agents which include three dimensions: quality attributes, pipelines, and artifacts. Each guardrail layer can be designed to protect specific quality attributes (such as privacy and security), specific pipeline stages (such as prompts, intermediate results and final results), as well as agent artifacts (such as goals, plans, and tools). While each layer may have its own weaknesses (i.e. holes in the Swiss Cheese Model), the combined layers create a robust defense against failures. This reference architecture provides concrete guidance for researchers and practitioners, enabling AI-safety-by-design from a software architecture perspective.

The rest of the paper is organized as follows. \ref{tax_related-works} discuss the related works. The research methods employed in this study are described in \ref{Methodology}. The proposed taxonomy of guardrails is presented in \ref{taxonomy}. \ref{sec-rq3-architecture} proposes the reference architecture for multi-layered runtime guardrails for FM-based agents. \ref{Threats to Validity} discusses threats, and \ref{Conclusion and Future Work} concludes the paper with future work.

\section{Background and Related Work}
\label{tax_related-works}
FMs have significantly advanced current agent development and emphasize the need to safeguard their behavior~\cite{qinghua-ref-architecture-2024, FM-survey-2023-bert}. In this context, guardrails for FM-based agents have been explored; however, but comprehensive studies on this topic are lacking. This paper addresses that gap. 

\subsection{Recent State-of-the-Art Works on Foundation Models and FM-Based Agents}
\label{Key Research on Foundation Models and FM-Based Agents}

In 2021, Bommasani et al.~\cite{29-ed-foundation-model-opportunity-risk2021} provided a comprehensive discussion on FMs, illustrating key elements, relationships, opportunities, and associated risks. While their focus was on FMs in general, they highlighted the potential for these models to serve as the foundation for more complex systems, including FM-based agents.
Zhou et al.~\cite{FM-survey-2023-bert} reviewed research advancements, challenges, and opportunities for pre-trained models in text, image, graph, and data modalities. They also discussed the integration of FMs into systems such as agents. Both works offer excellent insights into future research directions to address open problems and associated risks.

Recently, Lu et al. developed a taxonomy of FM-based systems focusing on their pre-training, adaptation, architectural design, and responsible-AI-by-design~\cite{lu-taxonomy-fms}. The taxonomy aids software architects and developers in evaluating and integrating FMs into complex agent systems. The authors then highlighted  considerations for responsible AI and safety attributes. Several other works~\cite{RAI-Future-Qinghua, RAI-book-qinghua, AI-risk-manage-2024,Guardagent-framework-2024,LLM-bsaed-autonomus-agenet-survey-2024} also emphasize the importance of responsible AI and safety  practices for FM-based agents. In~\cite{llm-state-risk-2,llm-current-state-2024}, the authors explored the risks associated with deploying LLM-based agents and evaluated current approaches for mitigating these risks through model alignment, respectively. 
In 2024, a reference architecture for designing responsible and safe FM-based agents is proposed in~\cite{qinghua-ref-architecture-2024}. The authors demonstrated that the unique characteristics of FM-based agents—such as their autonomous operation, non-deterministic behavior, and continuous evolution—pose significant challenges in ensuring responsible AI and AI safety.

\subsection{Guardrails Approaches and Tools for FM-Based Agents}
\label{tax-StateofArt-guardrails}
There exist several frameworks and tools for designing guardrails~\cite{13-silent-guardrial,33-ed-causal-guardrails,9-guardrails-for-chatboats,32-ed-Resilient-guardrails,25-red-teaming-to-improve-guardrails}. These works explored model alignment during design time to ensure that the FM's outputs align with defined goals. Pre-training and adaptation strategies play a significant role in mitigating risks in FM-based agents. Our focus, however, is on runtime guardrails that monitor and control the agent’s behavior during operation. These guardrails are essential for addressing emergent issues that arise during agent interactions within dynamic environments~\cite{29-ed-foundation-model-opportunity-risk2021,AI-risk-manage-2024}.

Some initial efforts have been made toward runtime guardrails. NeMo Guardrails~\cite{8-nemo-guardrails} provides programmable guardrails to ensure that agents operate within safe parameters by monitoring inputs and outputs. OpenAI's Moderation API~\cite{OpenAI's-Moderation-API} monitors and filters harmful content generated by agents to protect user interactions.
The GuardAgent framework~\cite{Guardagent-framework-2024} utilizes an agent to oversee and safeguard other agents. It demonstrates strong generalization and low operational overhead by dynamically generating guardrail code.
We found that continuous validation ensures outputs from FM-based agents adhere to predefined ethical standards and guidelines. Techniques such as auditing agents through multi-layered approaches~\cite{15-ed-auditing-llm, 18-ed-prompt-injection-llm} are used to check for biases and ensure ethical compliance. 

Recently, Bengio et. al.~\cite{AI-risk-manage-2024} demonstrate that adaptive fail-safes characteristics of guardrails intervene automatically when an FM-based agent exhibits potentially harmful behavior. These fail-safes are designed to modify or halt outputs that could lead to undesirable consequences. Similarly, dynamic access controls adjust access permissions in real time based on the context of data usage to protect sensitive information and ensure it is accessible under appropriate circumstances~\cite{tingthing-privacy}. Due to the dynamic and adaptive nature of
FM-based agents, designing effective runtime guardrails poses several additional challenges~\cite{RAI-book-qinghua,LLM-bsaed-autonomus-agenet-survey-2024} e.g.,  scalability of guardrail mechanisms, the need for real-time monitoring, and the complexity of interpreting agent behaviors in diverse contexts. The authors in~\cite{boming-ai-safety-taxonomy} propose a framework for evaluating AI systems, which is applicable to FM-based agents. It includes harmonized terminology, a taxonomy of key evaluation elements, and a mapping of the AI lifecycle to stakeholders for ethical and accountable deployment. Despite these efforts, no framework currently provides comprehensive guidance on designing multi-layered runtime guardrails for FM-based agents, which we explore in this paper based on SLR.

\section{Methodology}
\label{Methodology}
This study focuses on two primary concepts: (i) FM-based agents and (ii) multi-layered runtime guardrails. We adopted the Petticrew and Roberts approach~\cite{petticrew-systematic-picoc} to define the \textbf{P}opulation, \textbf{I}nterventions, \textbf{C}omparison, \textbf{O}utcomes, and \textbf{C}ontext (PICOC), within which the intervention in this study is delivered. The PICOC for this study is shown in~\ref{SLR_PICOC}. Using these PICOC components and following Kitchenham's guidelines~\cite{kitchenham-guideline-for-slr}, we developed the protocol for this study. 

\subsection{Research Approach}
\label{Research Scope and Protocol Development}
\begin{figure}[t]
  \centering
\hspace*{-.3in}\includegraphics[width=.575\textwidth]{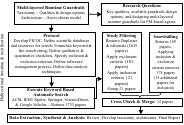}
  \caption[]{Methodology}
  \vspace{-.1in}
  \label{SLR_WorkFlow}

\end{figure}

The research approach for this study is summarized in~\ref{SLR_WorkFlow}. Initially, we determined the research scope and developed a protocol following Kitchenham's guidelines~\cite{kitchenham-guideline-for-slr}. 
The protocol guided the entire study by defining relevant scientific resources, formulating keywords and search strings, outlining qualitative and quantitative checklists, and specifying criteria for the inclusion and exclusion of studies\footnote{A detailed review protocol discussion is beyond this paper's scope; see supplementary materials for details~\cite{ICSA2025-Suplimental-GITHub-Shamsujjoha}.}. 

\clearpage

\subsection{Research Questions}
\label{Research Questions}
When formulating our Research Questions (RQ), we wanted to ensure that they were broad enough to capture the diverse aspects of multi-layered runtime guardrails while being specific enough to provide actionable insights. We captured these aspects through the following three RQs: 
\\
\\
\textbf{RQ1:} \textbf{What are essential qualities for designing runtime guardrails in FM-based agents?} 
\\
Our first research question studies the key qualities for designing multi-layered runtime guardrails in FM-based agents. \ref{tax-rq1} elaborate on how it is addressed. 
\\
\\
\textbf{RQ2:} \textbf{What are the design options for runtime guardrails in FM-based agents?}
\\
Our second research question investigates guardrails design options in FM-based agents from different perspectives, including action, target, scope, rule, autonomy, modalities, and underlying techniques. \ref{tax-rq2} outlines our approach to addressing this research question.
\\
\\
\textbf{RQ3:} \textbf{How can we design runtime guardrails to address the unique challenges of FM-based agents?}
\\
Our third research question explores how to address the challenges arising from the autonomous and non-deterministic nature of FM-based agents. Specially, we examine how to adapt the Swiss Cheese Model to safeguard the behaviors of FM-based agents by implementing multi-layered guardrails across various agent artifacts. \ref{sec-rq3-architecture} presents the proposed architecture and discusses our strategies for addressing this research question.

\begin{table}[!t]
\centering
\caption{PICOC for this Study}
\label{SLR_PICOC}  
\hspace*{.2in}\begin{tabular}{|l||p{.3\textwidth}|}
\hline
\textbf{Population} & Studies and researches focus on multi-layered runtime guardrails within foundation model-based agents. \\ \hline

\textbf{Intervention} & Development, optimization, and evaluation of multilayer runtime guardrails in foundation model-based agents, focusing on key quality attributes and design strategies similar to the Swiss Cheese Model structure.\\
\hline

\textbf{Comparison} & Comparative analysis of  approaches to design multi-layered runtime guardrails in FM-based agents.\\
\hline

\textbf{Outcomes} & Taxonomy of multi-layered runtime guardrails for foundation model-based agents.\\
\hline

\textbf{Context} &
\textbf{Include:} Empirical and theoretical studies on the components, design and evaluation of guardrails in foundation model-based agents.

\textbf{Exclude:} Studies beyond the scope of foundation model based agents, non-English literature, and those not considering guardrails.
\\\hline
\end{tabular}
\end{table}

\subsection{Study Search and Filtering Process}
\label{Study Search and Filtering Process}

To collect potentially relevant studies, we executed an automated keyword-based search and carried out the study filtering process in  March 2024. Consequently, studies published after this period were not investigated in this research. 
Our filtration process involved initial screening, grouping papers, and resolving conflicts in study selection. We also conducted manual searching and backward snowballing to ensure no significant studies were missed. The study filtering process, including the list of selected studies and their quality assessment scores, is detailed in supplementary materials~\cite{ICSA2025-Suplimental-GITHub-Shamsujjoha}. 
We have also made our study filtration file (`\textit{Study Filtration.xlsx}') publicly available in the supplementary materials. This file presents a step-by-step breakdown of the filtration process, narrowing the initially retrieved studies to the final selected studies. 
Once all relevant studies were cross-checked, identified, and collected, we performed data extraction, synthesis, and analysis. Finally, findings from 32 high-quality selected studies were used to develop the taxonomy and presented in this report.

\subsection{Data Extraction and Quality Assessment}
\label{Data Extraction and Quality Assessment}

We used a semi-automated process~\cite{data-extraction-2021-for-review} for data extraction from the selected studies to answer our RQs. Key qualitative information extracted from each selected study includes guardrails definitions, key quality attributes, and design options. We also extracted several relevant pieces of information to understand the context and considerations in designing and evaluating multi-layered runtime guardrails\footnote{Data extraction sheet is provided in the supplementary materials~\cite{ICSA2025-Suplimental-GITHub-Shamsujjoha}}. 

We  evaluated each study based on following five \textbf{Q}uality \textbf{A}ssessment \textbf{C}riteria (QAC) on a scale from 1 (Very Poor) to 5 (Excellent): (i) relevance to guardrails for FM-based agents, (ii) clarity of methodology for guardrail design, (iii) adequacy in data collection, analysis, and evaluation of guardrail effectiveness across different layers of the agent architecture, (iv) discussion of challenges in designing guardrails for autonomous and non-deterministic behaviors in agents, and (v) practical applicability of findings for guardrails in FM-based agents. If a study's average score was less than 2, it was excluded from further analysis. Otherwise, we used the qualitative information to decide this.

\section{Taxonomy of Guardrails for FM-based Agents}
\label{taxonomy}
\begin{figure}
\centering
\includegraphics[width=0.535\textwidth]{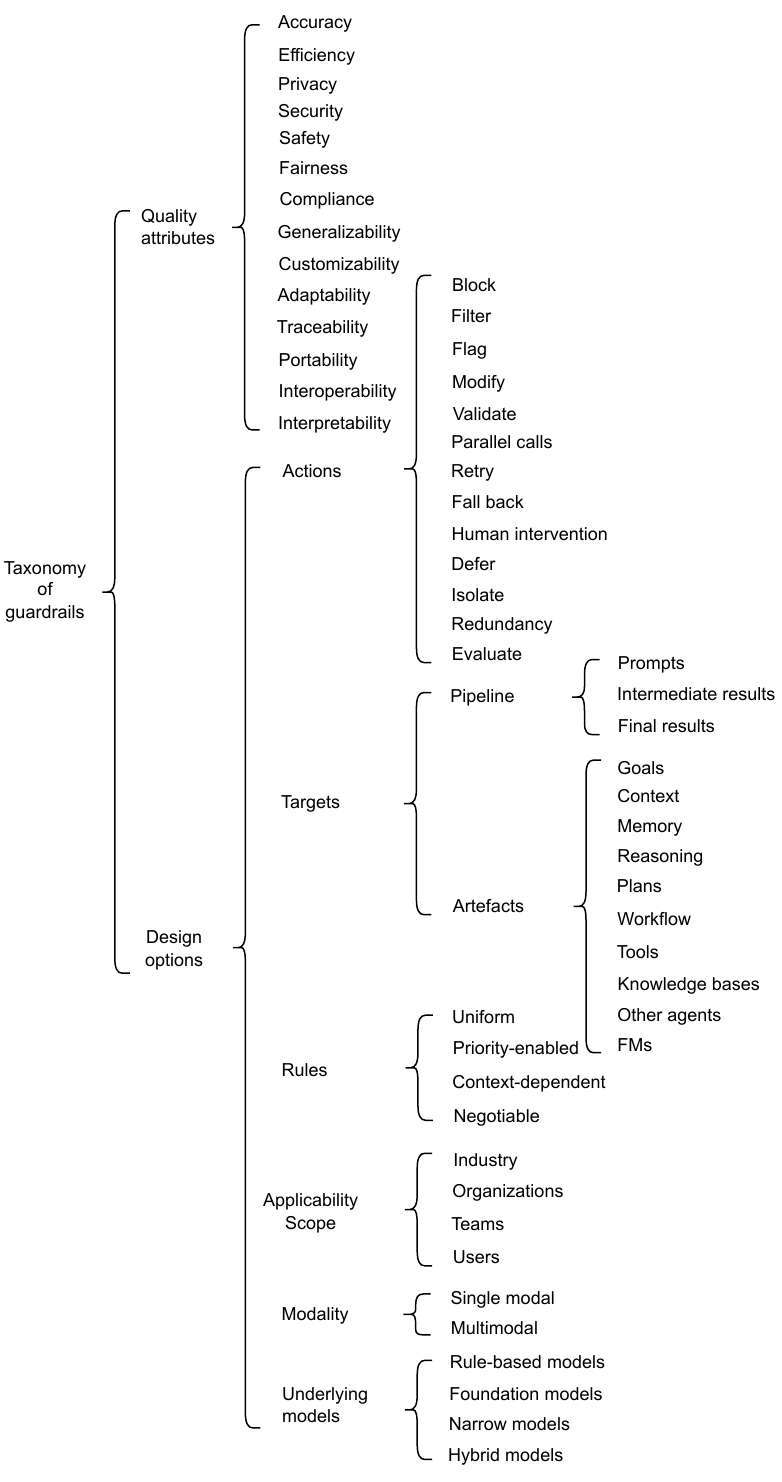}
\caption{Taxonomy of multi-layered runtime guardrails for FM-based agents.} \label{guardrails-tax-structure}
\vspace{-.2in}
\end{figure}

\ref{guardrails-tax-structure} presents the proposed taxonomy of runtime guardrails for FM-based agents, developed based on the results of a systematic literature review.  
The taxonomy was developed by synthesizing findings from 32 selected studies through an iterative clustering and validation process. It is organized into two key components: (i) Quality attributes, capturing relevant facets that influence guardrails’ effectiveness and reliability, and (ii) Design dimensions, representing practical approaches for implementing guardrails. The taxonomy combines external and internal quality attributes to merge overlapping categories, as some attributes (e.g., security, adaptability) apply to both developer-focused and stakeholder-facing contexts. Here, we also standardize terminology and refine overlapping attributes based on established practices in software engineering~\cite{kitchenham-guideline-for-slr,kitchenham2019problems}, while acknowledging interdependencies. This organization helps researchers and developers to distinguish between guardrails’ core characteristics from practical implementation approaches.

\subsection{Quality Attributes of Guardrails}
\label{tax-rq1}

We examine the key quality attributes essential for designing runtime guardrails. These attributes ensure guardrails meet critical performance, security, and reliability goals aligned with design objectives, while also improving end-user trust and system-level outcomes. They were systematically synthesized, clustered, and refined through an iterative process to emphasize their unique contributions. Below, we discuss these attributes in detail.

\subsubsection{\textbf{Accuracy}}
\label{tax-rq1-accuracy}
Accuracy in FM-based agents is crucial, particularly in mitigating issues such as hallucinations, misinformation, and disinformation~\cite{llm-hallucination-survey}. Hallucinations occur when models generate information that is factually incorrect. Such inaccuracies can mislead users and damage the credibility of the agent~\cite{28-ed-decoding-trust-gpt}. Misinformation refers to the unintentional spread of false information, while disinformation involves the deliberate dissemination of falsehoods to deceive users~\cite{FM-survey-2023-bert}.  
FM-based agents can incur significant costs due to errors, inaccuracies (leading to inefficiency), or non-compliance with regulations. Without proper guardrails, agents might generate outputs that lead to financial losses, legal penalties, or damage to their reputation~\cite{17-fontier-ai-risk,5-frontier-AI-risk}. 
Thus, leading companies e.g., OpenAI and Google use guardrails to clearly label AI-generated content to prevent deepfakes and misinformation~\cite{Open-AI-guardrails-protocol,MIT-AI-work-for-us}. 

\subsubsection{\textbf{Efficiency}}
\label{tax-rq1-performance}
Efficiency is crucial in FM-based agents, as users expect fast, efficient responses~\cite{LLM-survey-evaluation}. Without guardrails, agents risk engaging in resource-intensive tasks that slow down response times~\cite{16-ed-safety-ethical-guardrails}. By dynamically managing resources across multiple layers, these guardrails prevent inefficiencies, such as endless loops, and filter irrelevant inputs, ensuring that agents focus on processing meaningful data~\cite{7-Shield-llm,8-nemo-guardrails,qinghua-ref-architecture-2024}.

\subsubsection{\textbf{Privacy}}
\label{tax-rq1-privacy}
Privacy in FM-based agents poses risks due to handling sensitive data, where data leakage might expose confidential information~\cite{5-frontier-AI-risk,29-ed-foundation-model-opportunity-risk2021}. This leakage can occur through direct responses or statistical inferences, or inadvertent revelations through model outputs. In 2023, a notable incident involved Samsung employees leaking proprietary information into ChatGPT, leading to Samsung banning ChatGPT~\cite{samsung-ban-ChatGPT}. 

\subsubsection{\textbf{Security}}
\label{tax-rq1-security}
Security in FM-based agents involves protecting them from malicious activities that could compromise their integrity and functionality~\cite{19-ed-mitigate-security-llm,AI-Guardrials,foundation-model-stanford}. For example, an FM-based agent could be targeted by hackers to manipulate data, producing incorrect or harmful outputs that affect decision-making processes~\cite{11-ed-AIGC-Guardrail-Scenarios-2024}. An incident reported in~\cite{Tay-Chatboat-2021} described how malicious users manipulated Microsoft's Tay chatbot to produce inappropriate (offensive) content, leading to its shutdown.
FM-based agents are also vulnerable to hacks that may breach data confidentiality~\cite{32-ed-Resilient-guardrails}. Even with authorized access, there is a risk of data misuse by third-party providers~\cite{16-ed-safety-ethical-guardrails}. Moreover, FM-based agents are prone to adversarial attacks, where specially designed queries extract sensitive information. 
Guardrails mitigate these risks by detecting and responding to real-time threats across various operational layers, safeguarding agent integrity~\cite{19-ed-mitigate-security-llm,28-ed-decoding-trust-gpt}, confidentiality~\cite{13-silent-guardrial,20-jailbreak-llm,24-ed-guard-unsafe-llm}, availability~\cite{6-Code-help-llm-guardrails,32-ed-Resilient-guardrails,15-ed-auditing-llm,17-fontier-ai-risk,23-ed-safety-assessing-llm} and performance~\cite{11-ed-AIGC-Guardrail-Scenarios-2024,29-ed-foundation-model-opportunity-risk2021}. 

\subsubsection{\textbf{Safety}}
\label{tax-rq1-safety}
FM-based agents face significant safety issues, particularly in generating harmful or misleading outputs. These issues can arise when models produce content that is inappropriate, offensive, or incorrect~\cite{qinghua-ref-architecture-2024}. These issues are critical in contexts where FM-based agents handle critical data like medical diagnosis or self-driving cars, where inaccurate outputs could have severe consequences~\cite{30-ed-socio-technical-safety-GenAI}. Additionally, there is a risk of generating questionable content, which can damage the credibility and acceptance of the agent~\cite{32-ed-Resilient-guardrails}.

\subsubsection{\textbf{Fairness}}
\label{tax-rq1-fairness}
FM-based agents can face bias and discrimination in model outputs. These biases can emerge from the training data, model algorithms, or deployment context~\cite{formalizing-fairness-in-ML,RAI-book-qinghua}. For instance, an agent used in recruitment for screening CVs might inadvertently favor candidates from certain demographics, cultures, and languages~\cite{17-fontier-ai-risk,LLM-bsaed-autonomus-agenet-survey-2024}, affecting credibility.

\subsubsection{\textbf{Compliance}}
\label{tax-rq1-compliance}
Compliance in FM-based agents involves adhering to legal and regulatory standards~\cite{8-nemo-guardrails,FM-survey-2023-bert}. These issues are critical because non-compliance can lead to legal penalties, reputational damage, and loss of user trust. Runtime guardrails reduce these risks by ensuring alignment with data protection regulations, industry standards, and guidelines through continuous monitoring at multiple levels~\cite{16-ed-safety-ethical-guardrails,17-fontier-ai-risk}. 
Additionally, these guardrails assist in automating compliance checks. They ensure that all aspects of the FM-based agent's operations align with the necessary legal and regulatory frameworks~\cite{25-red-teaming-to-improve-guardrails,Guardagent-framework-2024}, and better support internal audits and external reviews~\cite{5-frontier-AI-risk}. For example,
FM-based agents may unintentionally facilitate unauthorized use of generated content, making it vulnerable to duplication or improper distribution~\cite{17-fontier-ai-risk,13-silent-guardrial,28-ed-decoding-trust-gpt}. Guardrails operating in real time help mitigate these risks by detecting and restricting unauthorized access, ensuring better copyright protection~\cite{24-ed-guard-unsafe-llm}.  
Techniques such as watermarking and labeling are applied across different layers to ensure the ownership and compliance with licensing laws~\cite{28-ed-decoding-trust-gpt,29-ed-foundation-model-opportunity-risk2021}.

\subsubsection{\textbf{Generalizability}}
\label{tex-rq2-key-atr-Generalizability}
Generalizability in  guardrails for FM-based agents refers to their ability to function effectively in real-time across multiple layers and diverse scenarios without prior configurations~\cite{generalizability-ai-model-2023}. Such guardrails ensure that protective measures are not overly specific to a single use case but can adapt to various contexts and still perform reliably across layers.
The agents' ability to handle diverse linguistic, cultural, and operational contexts is essential to provide robust protection, resilience, and reliability and is ensured by the generalizability attribute~\cite{5-frontier-AI-risk,29-ed-foundation-model-opportunity-risk2021}. Guardrails that can extend their applicability to new domains without significant reconfiguration or degradation in performance, even during unexpected inputs or data types, are essential~\cite{3-ed-build-guardrails-llm,7-Shield-llm}. 

\subsubsection{\textbf{Customizability}}
\label{tex-rq2-key-atr-Customizability}
Customizable guardrails provide tailored protection that meets specific requirements and supports diverse operational needs in FM-based agents~\cite{29-ed-foundation-model-opportunity-risk2021,Customizing-fm-2024}. The multi-layered runtime approach allows for customization at different layers to enable fine-grained control over the agent's behavior during execution, such as adjustments and configurations that align with particular operational goals, data characteristics, and regulatory environments.
For example, a customer service chatbot can enable priorities for different  guardrails and adjust data handling based on the user's location and ensuring compliance with regulation. 
%

\subsubsection{\textbf{Adaptability}}
\label{tex-rq2-key-atr-adapt}
Adaptability in  guardrails is known as their capability to adjust and remain effective under varying conditions and data landscapes as context evolves~\cite{17-fontier-ai-risk,LLM-survey-evaluation}. This attribute ensures robust and continuous protection by dynamically responding to changes in input data, usage patterns, and emerging threats without manual reconfiguration~\cite{3-ed-build-guardrails-llm}.  For example, a customer service chatbot can automatically update its guardrails to detect and block new offensive terms during  interactions. This includes incorporating new knowledge and advancements in threat detection techniques~\cite{16-ed-safety-ethical-guardrails,29-ed-foundation-model-opportunity-risk2021}.

\subsubsection{\textbf{Traceability}}
\label{tex-rq2-key-atr-trace}
The traceability attribute of guardrails tracks and records the origins, processes, and decision paths, such as input and output of FMs, external tools, etc.~\cite{lu-taxonomy-fms}. It involves maintaining detailed logs and records that can be audited to understand how decisions are made. 
For example, in a customer service chatbot, traceability ensures that every recommendation can be traced back to the data sources and algorithms used. This provides a clear audit trail for transparency and accountability.
Traceability also aids in identifying the root causes of issues to enable timely and accurate troubleshooting and improvement~\cite{17-fontier-ai-risk}, and helps in maintaining user trust and regulatory requirements~\cite{8-nemo-guardrails,28-ed-decoding-trust-gpt}. Additionally, comprehensive documentation of data sources and model modifications better support effective auditing and compliance checking~\cite{5-frontier-AI-risk}.

\subsubsection{\textbf{Portability}}
\label{tex-rq2-key-atr-portability}
Portability in guardrails for FM-based agents refers to the ability of these protective measures to be easily adapted and applied across different FM-based agents~\cite{lu-taxonomy-fms}. Multiple layer runtime guardrails allow individual layers to be transferred and integrated into different agents with minimal adjustments in real time. 
This includes ensuring that they function consistently across various FM architectures and environments, thereby maintaining their effectiveness and integrity regardless of the underlying technologies~\cite{17-fontier-ai-risk}.
For example, the same guardrail can be applied for content moderation in both a customer service chatbot and a social media platform, regardless of their underlying technology.
The benefits of designing portable guardrails include compatibility across multiple programming languages and frameworks facilitate their integration into diverse technological stacks~\cite{24-ed-guard-unsafe-llm}. These capabilities ensure that the guardrails remain effective and operational as the agent evolves or migrates to new environments. Portable guardrails also support seamless updates and improve scalability to maintain high standards of security and compliance while adapting to new technological advancements within agents~\cite{8-nemo-guardrails}.

\subsubsection{\textbf{Interoperability}}
\label{tex-rq2-key-atr-Interoperability}
Interoperable guardrails work seamlessly across differing agents, technologies and interface effectively with various components and services within different agents~\cite{lu-taxonomy-fms}. They ensure that security, privacy, and compliance protocols can be applied consistently, even in heterogeneous environments that utilize varied software and hardware components, or diverse technological ecosystems~\cite{8-nemo-guardrails,33-ed-causal-guardrails}. Guardrails that interface with various APIs and data formats also enable smooth communication and operation across different agents~\cite{17-fontier-ai-risk}. For example, they enable a customer service copilot and internal support system to share data securely and consistently. This promotes cohesive and unified security management, reducing the complexity of maintaining multiple disparate protective measures~\cite{29-ed-foundation-model-opportunity-risk2021}, and better support collaborative efforts and data sharing~\cite{24-ed-guard-unsafe-llm}.

\subsubsection{\textbf{Interpretability}}
\label{tex-rq2-key-atr-Interpretability}
Interpretability refers to the clarity and transparency with which guardrails and protective measures operate. Interpretability allows better  inspection and understanding of each layer's function during execution. This allows users and stakeholders to understand how decisions are made and actions are taken by models. Thus increasing trust and  accountability~\cite{28-ed-decoding-trust-gpt,10-talking-abt-llm}. For example, a chatbot in healthcare, can explain why certain advice is given or restricted. Transparent guardrails better facilitate auditing and compliance~\cite{15-ed-auditing-llm}. They also help users to understand that actions taken by guardrails can be clearly understood and verified~\cite{13-silent-guardrial}. This is essential for identifying and correcting errors, as well as for ensuring that the agent's operations align with ethical and regulatory 
standards. 

\subsection{Design Options of Guardrails}
\label{tax-rq2}
This section presents a structured taxonomy for designing guardrails, focusing on identifying various design alternatives.

\subsubsection{\textbf{Actions}}
\label{Guardrails Actions}
Guardrail actions are crucial for addressing the specific needs of FM-based agent artifacts. We have identified the following guardrail actions that can be applied to FM-based agents:

\vspace{.35em} 
\begin{itemize}
 \setlength\itemsep{.25em} 
 
\item \textbf{Block:} The \textit{block} action prevents specific inputs (such as user prompts) or outputs (such as content generated by FMs) from being processed or sent by various components (such as FMs and tools) in FM-based agents~\cite{16-ed-safety-ethical-guardrails}. For example, the \textit{block} action can reject the user prompts containing harmful instructions, thus preventing undesired outcomes. 

\item \textbf{Filter:} The \textit{filter} action involves scanning and removing undesired or irrelevant content from the inputs or outputs of different components in FM-based agents~\cite{22-ed-persuasion-challenge-llm,21-ed-llm-safety-against-prompt}. For instance, a filter may remove any personal data contained in the user prompts or the output generated by FMs.

\item \textbf{Flag:} The \textit{flag} action is used to mark specific inputs, outputs, operations within FM-based agents~\cite{8-nemo-guardrails}. For example, unusual transactions requested by the FM-based agent can be flagged for human review to ensure they comply with organizational policies~\cite{29-ed-foundation-model-opportunity-risk2021,RAI-Future-Qinghua}.

\item \textbf{Modify:} The \textit{modify} action allows for the adjustment of inputs or outputs of various components in  FM-based agents to meet specific requirements or standards~\cite{9-guardrails-for-chatboats}. For example, the user prompts can be modified by adding more context and examples, making it easier for the FM to accurately interpret the user's intentions and provide more relevant responses. 

\item \textbf{Validate:} The \textit{validate} action checks agent artifacts against predefined criteria to ensure they meet specified requirements or standards~\cite{21-ed-llm-safety-against-prompt,17-fontier-ai-risk}. For example, the plan generated by FM-based agents should be validated, e.g., through external verifier~\cite{valmeekam2024llms}, to ensure it is compliant with  regulatory policies.

\item \textbf{Parallel calls:} The \textit{parallel calls} action can send multiple requests to the agent/component to improve responsiveness, e.g., a user can send a prompt to the agent or an external service multiple times at the same time and select the better response~\cite{8-nemo-guardrails,32-ed-Resilient-guardrails}.  

\item \textbf{Retry:} The \textit{retry} action involves attempting a request again after an initial failure or unsatisfactory result~\cite{12-ed-jailbroken}.

\item \textbf{Fall back:} When one step in the workflow cannot be executed successfully, the \textit{fall back} action  redirect to the previous step and state~\cite{12-ed-jailbroken,Guaranteed-Safe-AI,8-nemo-guardrails}. 

\item \textbf{Human intervention:} The \textit{human intervention} action requires humans to review and approve specific outputs or decisions~\cite{8-nemo-guardrails,32-ed-Resilient-guardrails,13-silent-guardrial}. For example, responses involving sensitive medical advice might be flagged for human approval before being communicated to users. 

\item \textbf{Defer:} The \textit{defer} action postpones the processing of a request or task until specific conditions are met or additional information is available~\cite{1-ED-Guardrails-auto-ambigous-statement}.

\item \textbf{Isolate:} The \textit{isolate} action involves segregating a specific entity (e.g., user) or component to prevent interaction with the agent~\cite{6-Code-help-llm-guardrails,4-attack-to-llm,19-ed-mitigate-security-llm}. For example, an agent might isolate a compromised narrow AI model suspected of being poisoned with malicious data in a sandbox environment, preventing potential harm to the agent.

\item \textbf{Redundancy:} The \textit{redundancy} action involves implementing backup processes or components to ensure continuity and reliability in case of failures~\cite{17-fontier-ai-risk,8-nemo-guardrails}. For example, two sensors can be deployed to detect context information for an agent.

\item \textbf{Evaluate:} The \textit{evaluate} action involves assessing the results~\cite{29-ed-foundation-model-opportunity-risk2021}. For instance, an agent might ask another agent to evaluate its intermediate or final results.


\end{itemize}

\begin{table}[t]
\centering
\caption{A Mapping of Agent Targets to Guardrail Actions}
\label{action-target-table}
\begin{tabular}{|p{.025\textwidth}|p{.08\textwidth}|p{.3\textwidth}|}
\hline
\textbf{Type} & \textbf{Targets} & \textbf{Guardrail Actions} \\ \hline

\multirow{3}{*}{\rotatebox[origin=c]{90}{Pipeline\hspace{.2in}}} 

&Prompts & Block, filter, flag, modify, parallel calls, retry, defer, evaluate \\  \cline{2-3}

&Intermediate results & Flag, human intervention, evaluate \\  \cline{2-3}

&Final results & Block, filter, flag, modify, retry, fall back, human intervention, evaluate \\  \hline

\multirow{9}{*}{\rotatebox[origin=c]{90}{Artifacts\hspace{1in}}} 

&Goals & Validate, block, flag, modify, human intervention, defer \\ \cline{2-3}

&Context & Block, filter, flag, modify, evaluate \\ \cline{2-3}

&Memory & Block, filter, flag, modify, retry, human intervention, isolate, evaluate \\ \cline{2-3}

&Reasoning & Flag, modify, validate, human intervention \\ \cline{2-3}

&Plans & Block, flag, modify, validate, retry, fall back, human intervention, defer \\ \cline{2-3}

&Workflows & Validate, parallel calls, retry, fall back, human intervention, defer, evaluate \\ \cline{2-3}

&Tools & Block, parallel calls, retry, fall back, human intervention, defer, evaluate \\ \cline{2-3}

&Knowledge bases & Block, filter, flag, modify, retry, isolate, evaluate, redundancy \\ \cline{2-3}

&Other agents & Block, flag, parallel calls, retry, fall back, human intervention, defer, isolate, evaluate \\ \cline{2-3}

&FMs & Block, filter, flag, modify, parallel calls, retry, fall back, human intervention, isolate, evaluate, redundancy \\ \hline
\end{tabular}
\end{table}

\subsubsection{\textbf{Targets}}
\label{Targets for Guardrails}

Guardrail actions can be applied to various targets across multiple layers, including both pipelines and artifacts. Some guardrails are applied to the entire pipeline (including prompts, intermediate results, and final results), while others focus on specific artifacts (e.g., goals, context, reasoning, plans, memory, tools, knowledge bases, other agents, FMs). \ref{action-target-table} summarizes agent targets and their corresponding guardrail actions. These combinations are derived from a synthesis of findings and analytical inferences based on selected studies. Moreover, they  serve as illustrative examples not exhaustive or definitional, and intended to provide practical insights for potential application.

\begin{itemize}

\item \textbf{Prompts:} Prompts are the initial user inputs or queries. Guardrails on prompts help ensure that user prompts are relevant, appropriate, formatted correctly, and easier for FMs to understand~\cite{18-ed-prompt-injection-llm,20-jailbreak-llm,21-ed-llm-safety-against-prompt}. 

\item \textbf{Intermediate Results:} Intermediate results are the outputs generated at various stages during the workflow generation of agents, before reaching the final outputs. By monitoring intermediate results, guardrails can detect anomalies or inaccuracies before they propagate to the final results.

\item \textbf{Final Results:} Final results are the end outputs generated by agents, which are delivered to users or downstream systems. Guardrails ensure that the final results meet user expectations and comply with regulations and standards.

\item \textbf{Goals:} Ensuring that agents' goals align with human values and do not deviate from the human's intended goals~\cite{8-nemo-guardrails,24-ed-guard-unsafe-llm}.

\item \textbf{Context:} Monitoring the context that agents collect to ensure it is relevant information and appropriate~\cite{Guardagent-framework-2024}. 

\item \textbf{Memory:} Managing the agents' memory to retain relevant data and discard outdated or irrelevant information, while also preventing memory poisoning~\cite{Guardagent-framework-2024,7-Shield-llm}. 

\item \textbf{Reasoning:} Checking whether the reasoning is sound~\cite{RAI-Future-Qinghua}. 

\item \textbf{Plans:} Ensuring the generated plans align with human goals~\cite{16-ed-safety-ethical-guardrails,RAI-Future-Qinghua}. 

\item \textbf{Workflows:} Managing the exceptions happened during runtime workflow execution~\cite{lu2015tail}.

\item \textbf{Tools:} Overseeing the proper use of tools by agents, including implementing access controls, restricting tool capabilities, and detecting potential vulnerabilities~\cite{24-ed-guard-unsafe-llm,Guardagent-framework-2024}. 

\item \textbf{Knowledge Bases:} Guardrails enforce stringent monitoring and validation of external knowledge bases, particularly in retrieval augmented generation scenarios~\cite{31-ed-kang2020model}. For example, they can prevent the retrieval of sensitive business data~\cite{RAG-attack-2024}.



\item \textbf{Other Agents:} Managing interactions between agents to ensure collaboration, prevent conflicts, and mitigate risks associated with malicious behaviors~\cite{RAI-Future-Qinghua,24-ed-guard-unsafe-llm}. 

\item \textbf{FMs:} Guardrails ensures the outputs generated by FMs are relevant, appropriate and safe. Also, guardrails oversee the utilization of FMs, preventing misuse and ensuring their application under appropriate conditions~\cite{29-ed-foundation-model-opportunity-risk2021,FM-survey-2023-bert}. 

\item \textbf{Execution Time:} Managing execution time is crucial for ensuring optimal performance and timely responses for FM-based agents. Guardrails can enforce execution cut-offs for operations exceeding predefined thresholds, ensuring that agents prioritize responsiveness while avoiding excessive resource usage. For example, in conversational agents, guardrails can halt long-running processes to provide users with timely answers rather than delaying responses indefinitely.

\end{itemize}

\subsubsection{\textbf{Rules}}
\label{Rules}
Guardrails rules can be configured in different ways: including uniform rules, priority-enabled rules, context-dependent rules, and negotiable rules. A uniform strategy applies the same set of guardrails consistently across all scenarios, ensuring simplicity and uniformity~\cite{33-ed-causal-guardrails}. It is particularly effective in environments with stable and well-understood risks. It largely reduces the complexity of managing diverse guardrails~\cite{13-silent-guardrial}. A priority-enabled strategy prioritizes certain guardrails based on the criticality and sensitivity of operations or data. 
Context-dependent strategies adjust the implementation of guardrails based on the system's specific operational context. This allows for dynamic adjustments to guardrails in response to changing conditions, user needs, and operational environments~\cite{24-ed-guard-unsafe-llm}. 
The negotiability of guardrails, categorized into hard and soft, defines the level of flexibility in enforcing rules. Soft guardrails allow adjustments based on context and situational demands, providing a balance between protection and operational flexibility~\cite{24-ed-guard-unsafe-llm}. In contrast, hard guardrails are rigid and non-negotiable, ensuring adherence to critical legal, ethical, or safety standards~\cite{5-frontier-AI-risk,30-ed-socio-technical-safety-GenAI}.

\subsubsection{\textbf{Applicability Scope}}
\label{Scopes}
The applicability scope of guardrails in FM-based agents ranges from industry regulations and standards to individual preferences. 
Industry-level regulations and standards provide the broader regulatory framework within which FM-based agents must operate. Guardrails designed to comply with these regulations guarantee that the system adheres to industry best practices and legal requirements~\cite{8-nemo-guardrails}. They facilitate simpler auditing and certification processes, ensuring the agent remains compliant with evolving regulatory landscapes. \pagebreak

At the organizational level, guardrails align with internal policies and procedures governing the operation and use of FM-based agents. This includes compliance with corporate governance, data protection policies, and ethical guidelines established by the organization~\cite{5-frontier-AI-risk}. Guardrails also ensure consistency and accountability across different departments and functions within the organization.

Team-level constraints focus on the technical and operational limitations defined by the development team. Guardrails at this level ensure that the agent functions efficiently within these constraints, such as computational and memory limits, while maintaining robustness and reliability~\cite{17-fontier-ai-risk}. They also ensure that the agent’s operations do not exceed predefined thresholds that could lead to performance degradation or security vulnerabilities.

From the user perspective, guardrails can reflect individual preferences and requirements. This involves adjusting the agent’s behavior based on user-defined settings to align outputs with both user expectations and ethical considerations. Incorporating user preferences into guardrails provides a personalized experience while maintaining safety and compliance~\cite{13-silent-guardrial,22-ed-persuasion-challenge-llm}. Such guardrails ensure that the system respects user autonomy and produces outputs that are relevant and acceptable.

\subsubsection{\textbf{Modality}}
\label{Modality}
The modality of guardrails refers to the types of data and interactions they manage. Guardrails can be designed for single modal or multimodal systems. Single modal systems operate with one type of data input or output, such as text, image, or audio. For instance, in text-based agents, guardrails focus on addressing issues like offensive language, misinformation, and data privacy~\cite{24-ed-guard-unsafe-llm}. In image-based agents, they may involve techniques for detecting explicit content or ensuring image quality standards~\cite{17-fontier-ai-risk}.

Multimodal guardrails address the combined risks of handling multiple data types. They synchronize protections across different data types, ensuring comprehensive security and compliance~\cite{13-silent-guardrial}. For example, a system that generates text based on image inputs must ensure accurate and ethical representation of the image content. This requires advanced cross-modal analysis and validation techniques to ensure the system operates reliably and ethically across all data types~\cite{32-ed-Resilient-guardrails}.

\subsubsection{\textbf{Underlying models}}
\label{Underlying Technique}
The underlying techniques of guardrails include rule-based, hybrid, and machine learning models, with each representing a distinct design option to meet specific requirements~\cite{qinghua-ref-architecture-2024,lu-taxonomy-fms}.
Rule-based models utilize predefined rules to monitor and control FM-based agents behavior. These models implement strict and deterministic guidelines that the agent must follow to ensure compliance with regulatory requirements for data access and processing~\cite{24-ed-guard-unsafe-llm}. They are particularly effective in environments where operational parameters are well-defined and stable. 
Rule-based models can be updated and are somewhat flexible. However, they may still struggle with unexpected scenarios, such as detecting novel AI-generated content that falls outside predefined rules. This reliance on static rules can limit their adaptability, and regular updates are needed~\cite{Guaranteed-Safe-AI,8-nemo-guardrails}.

In contrast, machine learning models dynamically adapt and improve guardrails based on new data and scenarios. These models can also learn from historical data and identify patterns that indicate potential risks or compliance issues~\cite{7-Shield-llm}. Machine learning models can be further classified into narrow models and FMs. Narrow models are specialized systems designed for specific tasks or domains. They require targeted guardrails to address domain-specific risks and compliance needs~\cite{3-ed-build-guardrails-llm}. FMs are large, general-purpose models that serve as the backbone for multiple applications and tasks. These models necessitate comprehensive and scalable guardrails to handle a wide range of risks and compliance issues across different applications~\cite{17-fontier-ai-risk}. Nevertheless, they can be computationally intensive and require substantial data for training.

Hybrid models integrate rule-based approaches with the adaptability of machine learning models to respond to new threats and evolving data patterns~\cite{32-ed-Resilient-guardrails}. For instance, Khorramrouz et al.\cite{2-ed-toxicity-palm2} demonstrate the use of the PaLM 2 framework to process user input and dynamically implement rule-based decisions. This framework tests the system's limits by iteratively generating toxic content to evaluate PaLM 2's safety guardrails. However, integrating hybrid models can increase system complexity and create additional challenges~\cite{32-ed-Resilient-guardrails}. 

\begin{figure*}
\centering
\includegraphics[width=0.65\textwidth]{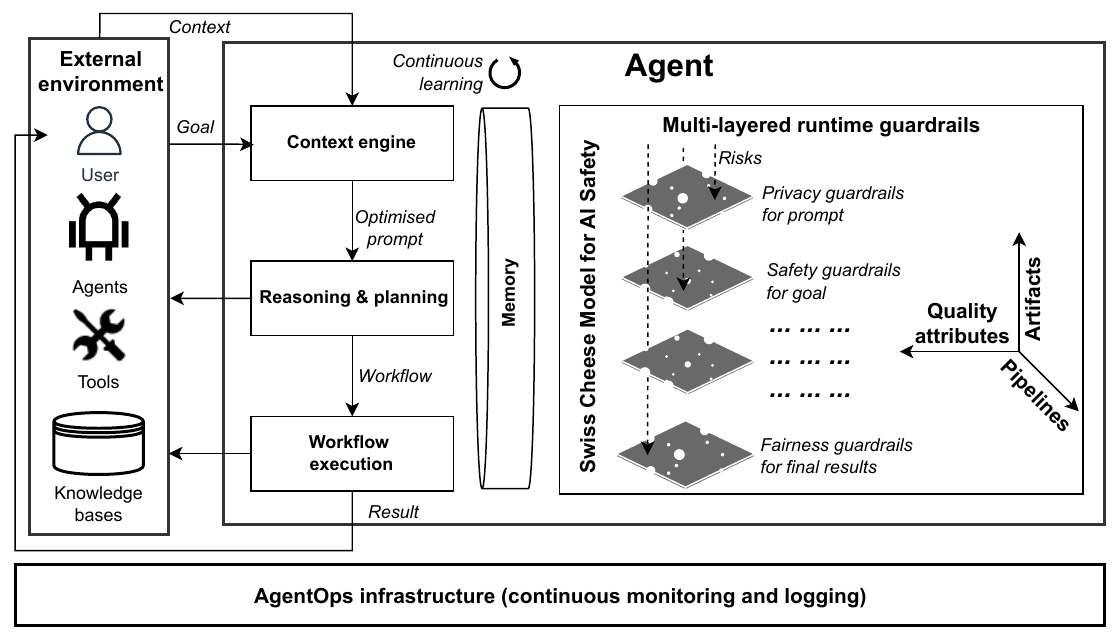}
\caption{Reference architecture for multi-layered guardrails of FM-based agents.} \label{overview}
\vspace{-2ex}
\end{figure*}

\section{Reference Architecture for Designing Multi-layered Runtime Guardrails of Agents} 
\label{sec-rq3-architecture}

 \ref{overview} illustrates the proposed reference architecture for multi-layered runtime guardrails of FM-based agents. It consists of four key components: (i) the external environment, (ii) agent components, (iii) built-in multi-layered runtime guardrails, and (iv) AgentOps infrastructure. This architecture is derived from our SLR findings (gaps and recommendations in the selected studies) and builds on empirical methodologies proposed in~\cite{Empirically-grounded-reference-architectures}. However, operational validation of the architecture lies beyond this paper’s scope. In the following sub-sections, we describe the components and mechanisms of this architecture and demonstrate how guardrails can be effectively implemented in FM-based agents.
 
\subsection{External Environment:}
\label{External Environments}
The external environment refers to all entities interacting with the agent at runtime, including users, other agents, external tools, and knowledge bases. Users provide goals and contextual inputs that shape the agent's objectives. To achieve user goals, the agent may utilize context detected in the external environment and interact with other agents, specialized tools, and extensive knowledge bases to perform complex tasks. 
 
\subsection{Agent Components}
\label{Agent Components}
Within the agent, there are four primary components: the context engine, reasoning and planning, workflow execution, and memory.

\begin{itemize}
    \item \textbf{Context Engine:} 
    The context engine processes multimodal context data from the external environment to enrich the user prompt, helping FMs better understand user goals. A prompt may contain elements such as goals and context. Instead of waiting for users' instructions, the agent can also proactively make suggestions based on the context it detects, such as screen recordings, eye tracking data, gestures, and document annotations~\cite{qinghua-ref-architecture-2024}. 

    \item \textbf{Reasoning and Planning:} After receiving optimized prompts, the reasoning and planning component processes the prompt to determine the most effective way of achieving the specified goal. This process may involve adopting reasoning patterns, such as the chain-of-thought pattern~\cite{wei2022chain}, which structures the agent's thinking into sequential, logical steps that align with the agent's objectives. A detailed plan is then formulated to outline each step required to accomplish the goal. This includes selecting the appropriate tools, knowledge bases, and agents to carry out each action. The memory component may be integrated to allow the agent to recall previously gathered experience and knowledge to refine the plan.

    \item \textbf{Workflow Execution:} The workflow execution component is responsible for executing the sequence of actions outlined by the reasoning and planning component. This component directly interacts with external tools, knowledge bases, and other agents to complete tasks and generate outputs aligned with the user's goals. The results are returned to the external environment and stored in the agent’s memory for future reference.

    \item \textbf{Memory:} The memory component in this architecture stores relevant information from prior interactions, plans, and results. This accumulated knowledge supports continuous learning, enabling the agent to refine its strategies and improve capabilities and skills over time, thereby improving accuracy and minimizing errors. 
\end{itemize}
 
\subsection{Multi-layered Runtime Guardrails}

\label{Multi-layered Runtime Guardrails}
Building on the Swiss Cheese Model, we design multi-layered runtime guardrails for FM-based agents, structured around the dimensions of quality attributes, pipelines, and artifacts  specified in the taxonomy. 
Here, we operationalize the target mappings (discussed earlier in \ref{Targets for Guardrails}) to illustrate their implementation within the proposed reference architecture.
In this architecture, each `cheese slice' represents a protective layer within the agent system, addressing quality attributes, pipeline stages, and/or specific artifacts, such as one concerning privacy guardrails for prompts or security guardrails for tools. While each layer contains holes (i.e., potential gaps or weaknesses), where risks might slip through, these holes are positioned differently across layers. Gaps in one layer are often covered by another; thus, even if one layer fails, another can catch and mitigate the issue.

From the perspective of \textbf{quality attributes} (discussed in \ref{tax-rq1}), guardrails can be designed to ensure accuracy, efficiency, privacy, security, safety, fairness, compliance. From the \textbf{pipelines} perspective, guardrails can be applied at multiple stages: the user prompts, intermediate results during workflow executions, and final results generated by the agent. 

\begin{itemize}
    \item \textbf{Guardrails for prompts:} Analyse incoming user prompts to detect and manage sensitive information, harmful content, misinformation, disinformation, discriminatory language, ensuring the prompt aligns with safety and ethical standards~\cite{liu2023prompt}. 

    \item \textbf{Guardrails for intermediate results:} Apply at each step of the workflow to verify that intermediate results are accurate, safe, and responsible, safeguarding the integrity of the process before the final results are produced. 

    \item \textbf{Guardrails for final results:} Check that the agent's final outputs are align with the user goals and governance requirements, such as AI safety standard requirements. 
\end{itemize}

Moreover, from the \textbf{artifacts} perspective, guardrails can be enforced on each agent artifact including goals, context, memory, reasoning, plans, tools, knowledge bases, other agents, and FMs. These guardrails ensure that each artifact is within safe and responsible boundaries. \pagebreak
\begin{itemize}
    \item \textbf{Guardrails for goals:} Ensure that the goals are achievable, within the agent's scope, and aligned with governance requirements, including regulatory standards and organizational policies, avoiding goals that may lead to harmful outcomes and potential misuse~\cite{yohsua2024international}. 
    
    \item \textbf{Guardrails for context:} Validate contextual information to ensure it is relevant, accurate, and free from sensitive or misleading information. 
    
    \item \textbf{Guardrails for memory:} Ensure that stored past experience is relevant, accurate, and free from any malicious or misleading content, preventing memory poisoning~\cite{chen2024agentpoison} and retaining only useful data for future interactions.
    
    \item \textbf{Guardrails for reasoning:} Check the agent's reasoning processes to prevent logical errors and ensure the reasoning steps are safe, responsible, and aligned with the user intent.
    
    \item \textbf{Guardrails for plans:} Assess the feasibility, safety, and compliance of the plans generated by the agent, ensuring that each step in the workflow is responsible and does not introduce unnecessary risk. The plan can be made by external verifiers, i.e., external planning tools~\cite{valmeekam2024llms}. 
    
    \item \textbf{Guardrails for workflows:} Handle the exceptions  during the workflow executions by implementing mechanisms like force-failing a step or retrying a tool call~\cite{lu2015tail}
    
    \item \textbf{Guardrails for external tools:} Analyse the quality (e.g. vulnerability~\cite{sun2023aspect}) of the external tools to ensure that only approved and safe tools are invoked by the agent. 
    \item \textbf{Guardrails for knowledge bases:} Verify that the information retrieved from knowledge bases is relevant and ethical (e.g., without any PII information). 
    \item \textbf{Guardrails for other agents:} Ensure the selected agents have a reliable and safe operational history.
    \item \textbf{Guardrails for FMs:} Enforce boundaries on FM’s non-deterministic outputs by applying modifications/flags 
\end{itemize}

\subsection{AgentOps}
\label{AgentOps}
AgentOps provides a comprehensive infrastructure designed to enable observability~\cite{dong2024taxonomy} for FM-based agents by continuously monitoring and recording runtime data. This infrastructure captures a wide range of data elements, from pipeline execution details and agent artifacts to the specific guardrails applied to the pipeline and artifacts. All these data need to be kept as evidence with metadata such as FM version and the timestamp. The data collected by the AgentOps infrastructure can also feed into multi-layered guardrails to activate the relevant guardrails as needed.

\section{Threats to Validity}
\label{Threats to Validity}
Our study is subject to standard literature search and selection bias threats. We addressed these threats by searching the most commonly used databases in the IT and software engineering domains. We revised  our search strings several times during the automatic search to maximize the number of relevant articles matching two key concepts: `guardrails' and `FM-based agents'. We also kept our search string generic to search through the titles, abstracts, keywords, and full text of articles to cover the maximum number of relevant papers. We then conducted a manual search on Google Scholar to complement the automatic search using a snowballing strategy. Furthermore, predefined review protocols with detailed inclusion and exclusion criteria helped us reduce bias in selecting primary studies. %
We applied several quality assessment criteria to estimate the quality of the selected primary studies. Even though the proposed criteria were not too strict,  applying them led to several initially selected papers being excluded. To mitigate the risk of missing important data from the primary studies, we reinstated the excluded papers that were closely related to the primary studies. 

Moreover, our definitions and categorizations may not capture all relevant aspects of guardrails in FM-based agents. To mitigate this threat, we validated the taxonomy through extensive literature review and expert feedback. However, this introduces a risk of producing biased results that address only experts' needs, as the people involved in the feedback process have extensive experience in the AI and software engineering domains. Our review protocols helped us to reduce such bias.

We prepared a guardrails taxonomy and conducted a comparative analysis of its components to help the reader better understand their design and evaluation. We critically examined the strength and consistency of relationships in the selected studies to develop a reliable taxonomy and reference architecture for designing built-in multi-layered runtime guardrails. Finally, we draw conclusions; nonetheless, the generalizability of guardrails to different contexts and types in FM-based agents remains a potential limitation. Specific adaptations might be necessary for certain systems, such as those used in healthcare or financial organizations. Additionally, the rapid evolution of FM-based agents and their associated guardrails may lead to parts of our findings becoming outdated. To address this, continual re-evaluation and refinement of the proposed taxonomy and reference architecture will be necessary over time.

\section{Conclusion and Future Work}
\label{Conclusion and Future Work}

This paper presents a comprehensive taxonomy of guardrails, derived from the results of an SLR, to advance the understanding of runtime guardrails for FM-based agents.
The taxonomy identifies 14 essential qualities for designing runtime guardrails in FM-based agents, addressing RQ1 in \ref{tax-rq1}. Subsequently, we outline design options for runtime guardrails, including actions, targets, employed rules, applicability scope, modality, and underlying models, addressing RQ2 in \ref{tax-rq2}.
We then propose a multi-layered runtime guardrail architecture based on the taxonomy, SLR findings, and the empirically grounded reference architecture approach~\cite{Empirically-grounded-reference-architectures}. This novel guardrail architecture better addresses the unique challenges associated with FM-based agents, as discussed in \ref{sec-rq3-architecture}, thus directly answering RQ3.
In future work, we plan to develop guardrail services for a scientific agent platform. These services will implement the proposed reference architecture and integrate the various design options outlined in the taxonomy.

 %


\begin{thebibliography}{10}
\providecommand{\url}[1]{#1}
\csname url@samestyle\endcsname
\providecommand{\newblock}{\relax}
\providecommand{\bibinfo}[2]{#2}
\providecommand{\BIBentrySTDinterwordspacing}{\spaceskip=0pt\relax}
\providecommand{\BIBentryALTinterwordstretchfactor}{4}
\providecommand{\BIBentryALTinterwordspacing}{\spaceskip=\fontdimen2\font plus
\BIBentryALTinterwordstretchfactor\fontdimen3\font minus \fontdimen4\font\relax}
\providecommand{\BIBforeignlanguage}[2]{{%
\expandafter\ifx\csname l@#1\endcsname\relax
\typeout{** WARNING: IEEEtran.bst: No hyphenation pattern has been}%
\typeout{** loaded for the language `#1'. Using the pattern for}%
\typeout{** the default language instead.}%
\else
\language=\csname l@#1\endcsname
\fi
#2}}
\providecommand{\BIBdecl}{\relax}
\BIBdecl

\bibitem{29-ed-foundation-model-opportunity-risk2021}
R.~Bommasani, D.~A. Hudson, E.~Adeli, and et~al., ``On the opportunities and risks of foundation models,'' \emph{CoRR}, vol. abs/2108.07258, 2021.

\bibitem{qinghua-ref-architecture-2024}
Q.~Lu, L.~Zhu, X.~Xu, Z.~Xing, S.~Harrer, and J.~Whittle, ``Towards responsible generative {AI:} a reference architecture for designing foundation model based agents,'' in \emph{21st International Conference on Software Architecture Companion}.\hskip 1em plus 0.5em minus 0.4em\relax IEEE, 2024, pp. 119--126.

\bibitem{RAI-book-qinghua}
Q.~Lu, L.~Zhu, J.~Whittle, and X.~Xu, \emph{Responsible {AI:} Best practices for creating trustworthy {AI} systems}, 1st~ed.\hskip 1em plus 0.5em minus 0.4em\relax Addison-Wesley Professional, 2023.

\bibitem{bass25engineering}
L.~Bass, Q.~Lu, I.~Weber, and L.~Zhu, \emph{Engineering {AI} Systems: Architecture and DevOps Essentials}.\hskip 1em plus 0.5em minus 0.4em\relax Addison-Wesley, 2025.

\bibitem{boming-ai-safety-taxonomy}
B.~Xia, Q.~Lu, L.~Zhu, and Z.~Xing, ``Towards {AI} safety: A taxonomy for {AI} system evaluation,'' \emph{arXiv preprint arXiv:2404.05388}, 2024.

\bibitem{9-guardrails-for-chatboats}
Y.~Wang and L.~Singh, ``Adding guardrails to advanced chatbots,'' \emph{arXiv preprint arXiv:2306.07500}, 2023.

\bibitem{28-ed-decoding-trust-gpt}
B.~Wang, W.~Chen, H.~Pei, C.~Xie, M.~Kang, C.~Zhang, C.~Xu, Z.~Xiong, R.~Dutta, R.~Schaeffer \emph{et~al.}, ``{DecodingTrust}: A comprehensive assessment of trustworthiness in {GPT} models,'' in \emph{Advances in Neural Information Processing Systems}, 2023, pp. 1--110.

\bibitem{23-ed-safety-assessing-llm}
B.~Wei, K.~Huang, Y.~Huang, T.~Xie, X.~Qi, M.~Xia, P.~Mittal, M.~Wang, and P.~Henderson, ``Assessing the brittleness of safety alignment via pruning and low-rank modifications,'' \emph{arXiv preprint arXiv:2402.05162}, 2024.

\bibitem{5-frontier-AI-risk}
M.~Anderljung \emph{et~al.}, ``Frontier {AI} regulation: Managing emerging risks to public safety,'' \emph{arXiv preprint arXiv:2307.03718}, 2023.

\bibitem{12-ed-jailbroken}
A.~Wei, N.~Haghtalab, and J.~Steinhardt, ``Jailbroken: How does {LLM} safety training fail?'' \emph{Advances in Neural Information Processing Systems}, vol.~36, 2024.

\bibitem{AI-Guardrials}
\BIBentryALTinterwordspacing
A.~Gubkin, ``Understanding why {AI} guardrails are necessary: Ensuring ethical and responsible {AI} use,'' 2024, {Last accessed on Jul.-2024}. [Online]. Available: \url{https://www.aporia.com/learn/ai-guardrails/}
\BIBentrySTDinterwordspacing

\bibitem{3-ed-build-guardrails-llm}
Y.~Dong, R.~Mu, G.~Jin, Y.~Qi, J.~Hu, X.~Zhao, J.~Meng, W.~Ruan, and X.~Huang, ``Building guardrails for large language models,'' \emph{arXiv preprint arXiv:2402.01822}, 2024.

\bibitem{8-nemo-guardrails}
T.~Rebedea, R.~Dinu, M.~N. Sreedhar, C.~Parisien, and J.~Cohen, ``{N}e{M}o guardrails: A toolkit for controllable and safe {LLM} applications with programmable rails,'' in \emph{Proceedings of the 2023 Conference on Empirical Methods in Natural Language Processing: System Demonstrations}, Dec. 2023, pp. 431--445.

\bibitem{31-ed-kang2020model}
D.~Kang, D.~Raghavan, P.~Bailis, and M.~Zaharia, ``Model assertions for monitoring and improving {ML} models,'' \emph{Proceedings of Machine Learning and Systems}, vol.~2, pp. 481--496, 2020.

\bibitem{15-ed-auditing-llm}
J.~M{\"o}kander, J.~Schuett, H.~R. Kirk, and L.~Floridi, ``Auditing large language models: A three-layered approach,'' \emph{AI and Ethics}, pp. 1--31, 2023.

\bibitem{19-ed-mitigate-security-llm}
A.~Kumar, S.~Singh, S.~V. Murty, and S.~Ragupathy, ``The ethics of interaction: Mitigating security threats in {LLMs},'' \emph{arXiv preprint arXiv:2401.12273}, 2024.

\bibitem{FM-survey-2023-bert}
C.~Zhou \emph{et~al.}, ``A comprehensive survey on pretrained foundation models: A history from {BERT to ChatGPT},'' \emph{arXiv preprint arXiv:2302.09419}, 2023.

\bibitem{swiss-cheesem-risk-management-review}
T.~Shabani, S.~Jerie, and T.~Shabani, ``A comprehensive review of the swiss cheese model in risk management,'' \emph{Safety in Extreme Environments}, vol.~6, no.~1, pp. 43--57, 2024.

\bibitem{lu-taxonomy-fms}
Q.~Lu, L.~Zhu, X.~Xu, Y.~Liu, Z.~Xing, and J.~Whittle, ``A taxonomy of foundation model based systems through the lens of software architecture,'' in \emph{Proceedings of the IEEE/ACM 3rd International Conference on {AI} Engineering-Software Engineering for AI}, 2024, pp. 1--6.

\bibitem{RAI-Future-Qinghua}
Q.~Lu, L.~Zhu, X.~Xu, Z.~Xing, S.~Harrer, and J.~Whittle, ``Building the future of responsible {AI}: A reference architecture for designing large language model based agents,'' \emph{arXiv e-prints}, 2023.

\bibitem{AI-risk-manage-2024}
Y.~Bengio \emph{et~al.}, ``Managing extreme {AI} risks amid rapid progress,'' \emph{Science}, vol. 384, no. 6698, pp. 842--845, 2024.

\bibitem{Guardagent-framework-2024}
Z.~Xiang \emph{et~al.}, ``{GuardAgent}: Safeguard llm agents by a guard agent via knowledge-enabled reasoning,'' \emph{arXiv preprint arXiv:2406.09187}, 2024.

\bibitem{LLM-bsaed-autonomus-agenet-survey-2024}
L.~Wang \emph{et~al.}, ``A survey on large language model based autonomous agents,'' \emph{Frontiers of Computer Science}, vol.~18, no.~6, pp. 1--26, 2024.

\bibitem{llm-state-risk-2}
X.~Tang, Q.~Jin, K.~Zhu, T.~Yuan, Y.~Zhang, W.~Zhou, M.~Qu, Y.~Zhao, J.~Tang, Z.~Zhang \emph{et~al.}, ``Prioritizing safeguarding over autonomy: Risks of {LLM} agents for science,'' \emph{arXiv preprint arXiv:2402.04247}, 2024.

\bibitem{llm-current-state-2024}
S.~G. Ayyamperumal and L.~Ge, ``Current state of {LLM} risks and {AI} guardrails,'' \emph{arXiv preprint arXiv:2406.12934}, 2024.

\bibitem{13-silent-guardrial}
J.~Zhao, K.~Chen, X.~Yuan, Y.~Qi, W.~Zhang, and N.~Yu, ``Silent guardian: Protecting text from malicious exploitation by large language models,'' \emph{arXiv preprint arXiv:2312.09669}, 2023.

\bibitem{33-ed-causal-guardrails}
Z.~Chu, Y.~Wang, L.~Li, Z.~Wang, Z.~Qin, and K.~Ren, ``A causal explainable guardrails for large language models,'' \emph{arXiv preprint arXiv:2405.04160}, 2024.

\bibitem{32-ed-Resilient-guardrails}
Z.~Yuan, Z.~Xiong, Y.~Zeng, N.~Yu, R.~Jia, D.~Song, and B.~Li, ``{RigorLLM}: Resilient guardrails for large language models against undesired content,'' \emph{arXiv preprint arXiv:2403.13031}, 2024.

\bibitem{25-red-teaming-to-improve-guardrails}
R.~R. Llaca, V.~Leskoschek, V.~C. Paiva, C.~Lup{\u{a}}u, P.~Lippmann, and J.~Yang, ``Student-teacher prompting for red teaming to improve guardrails,'' in \emph{Proceedings of the ART of Safety: Workshop on Adversarial testing and Red-Teaming for generative AI}, 2023, pp. 11--23.

\bibitem{OpenAI's-Moderation-API}
\BIBentryALTinterwordspacing
OpenAI, ``{OpenAI's} moderation {API},'' 2024, {Last accessed on Jul.-2024}. [Online]. Available: \url{https://platform.openai.com/docs/guides/moderation/overview}
\BIBentrySTDinterwordspacing

\bibitem{18-ed-prompt-injection-llm}
P.~Rai, S.~Sood, V.~K. Madisetti, and A.~Bahga, ``Guardian: A multi-tiered defense architecture for thwarting prompt injection attacks on {LLM}s,'' \emph{Journal of Software Engineering and Applications}, vol.~17, no.~1, pp. 43--68, 2024.

\bibitem{tingthing-privacy}
T.~Bi, G.~Yu, Q.~Lu, X.~Xu, and N.~Van~Beest, ``The privacy pillar - {A} conceptual framework for foundation model-based systems,'' \emph{arXiv preprint arXiv:2311.06998}, 2023.

\bibitem{petticrew-systematic-picoc}
M.~Petticrew and H.~Roberts, \emph{Systematic reviews in the social sciences: A practical guide}.\hskip 1em plus 0.5em minus 0.4em\relax John Wiley \& Sons, 2008.

\bibitem{kitchenham-guideline-for-slr}
B.~A. Kitchenham, S.~Charters, and {Other Keele Staffs}, ``Guidelines for performing systematic literature reviews in software engineering (version 2.3),'' Keele University and Durham University Joint Report, Tech. Rep., 2007.

\bibitem{ICSA2025-Suplimental-GITHub-Shamsujjoha}
M.~Shamsujjoha, Q.~Lu, D.~Zhao, and L.~Zhu, ``Supplementary materials: Systematic literature review protocol, study filtration sheet, and data extraction sheet for this paper,'' 2025, {Available at} \url{https://github.com/dishacse/Publication-Resources/tree/main/2025%20ICSA} Accessed: Jan-2025.

\bibitem{data-extraction-2021-for-review}
L.~Schmidt, A.~Finnerty~Mutlu, R.~Elmore, B.~Olorisade, J.~Thomas, and J.~Higgins, ``Data extraction methods for systematic review (semi)automation: Update of a living systematic review,'' \emph{F1000Research}, vol.~10, no. 401, 2023.

\bibitem{kitchenham2019problems}
\BIBentryALTinterwordspacing
B.~Kitchenham, L.~Madeyski, and P.~Brereton, ``Problems with statistical practice in human-centric software engineering experiments,'' in \emph{Proceedings of the Evaluation and Assessment on Software Engineering}.\hskip 1em plus 0.5em minus 0.4em\relax ACM, 2019, p. 134–143. [Online]. Available: \url{https://doi.org/10.1145/3319008.3319009}
\BIBentrySTDinterwordspacing

\bibitem{llm-hallucination-survey}
V.~Rawte, A.~Sheth, and A.~Das, ``A survey of hallucination in large foundation models,'' \emph{arXiv preprint arXiv:2309.05922}, 2023.

\bibitem{17-fontier-ai-risk}
S.~Ee, J.~O’Brien, Z.~Williams, A.~El-Dakhakhni, M.~Aird, and A.~Lintz, ``Adapting cybersecurity frameworks to manage frontier {AI} risks,'' Institute for AI Policy and Strategy, Tech. Rep., 2023.

\bibitem{Open-AI-guardrails-protocol}
\BIBentryALTinterwordspacing
OpenAI, ``{OpenAI} safety update,'' 2024. [Online]. Available: \url{https://openai.com/index/openai-safety-update/}
\BIBentrySTDinterwordspacing

\bibitem{MIT-AI-work-for-us}
\BIBentryALTinterwordspacing
G.~F. Marcus, \emph{Taming Silicon Valley: How We Can Ensure That AI Works for Us}.\hskip 1em plus 0.5em minus 0.4em\relax MIT Press, 2024. [Online]. Available: \url{https://mitpress.mit.edu/9780262551069/taming-silicon-valley/}
\BIBentrySTDinterwordspacing

\bibitem{LLM-survey-evaluation}
Y.~Chang, X.~Wang, J.~Wang, Y.~Wu, L.~Yang, K.~Zhu, H.~Chen, X.~Yi, C.~Wang, Y.~Wang, W.~Ye, Y.~Zhang, Y.~Chang, P.~S. Yu, Q.~Yang, and X.~Xie, ``A survey on evaluation of large language models,'' \emph{ACM Trans. Intell. Syst. Technol.}, vol.~15, no.~3, mar 2024.

\bibitem{16-ed-safety-ethical-guardrails}
S.~Banerjee, S.~Layek, R.~Hazra, and A.~Mukherjee, ``How (un) ethical are instruction-centric responses of {LLMs}? unveiling the vulnerabilities of safety guardrails to harmful queries,'' \emph{arXiv preprint arXiv:2402.15302}, 2024.

\bibitem{7-Shield-llm}
Z.~Zhang, Y.~Lu, J.~Ma, D.~Zhang, R.~Li, P.~Ke, H.~Sun, L.~Sha, Z.~Sui \emph{et~al.}, ``{ShieldLm}: Empowering {LLMs} as aligned, customizable and explainable safety detectors,'' \emph{arXiv preprint arXiv:2402.16444}, 2024.

\bibitem{samsung-ban-ChatGPT}
\BIBentryALTinterwordspacing
S.~Ray, ``Samsung bans {ChatGPT} among employees after sensitive code leak,'' 2023, {News article, Last accessed on Jul.-2024}. [Online]. Available: \url{https://www.forbes.com/sites/siladityaray/2023/05/02/samsung-bans-chatgpt-and-other-chatbots-for-employees-after-sensitive-code-leak/}
\BIBentrySTDinterwordspacing

\bibitem{foundation-model-stanford}
\BIBentryALTinterwordspacing
R.~Bommasani and P.~Liang, ``Reflections on foundation models,'' Stanford, Tech. Rep., 2021, {Last accessed on Jun.-2024}. [Online]. Available: \url{https://hai.stanford.edu/news/reflections-foundation-models}
\BIBentrySTDinterwordspacing

\bibitem{11-ed-AIGC-Guardrail-Scenarios-2024}
W.~Du, Q.~Li, J.~Zhou, X.~Ding, X.~Wang, Z.~Zhou, and J.~Liu, ``Finguard: A multimodal {AIGC} guardrail in financial scenarios,'' in \emph{Proceedings of the 5th ACM International Conference on Multimedia in Asia}, Taiwan, 2024, pp. 1--3.

\bibitem{Tay-Chatboat-2021}
T.~Zem{\v{c}}{\'\i}k, ``Failure of chatbot {Tay} was evil, ugliness and uselessness in its nature or do we judge it through cognitive shortcuts and biases?'' \emph{AI \& SOCIETY}, vol.~36, pp. 361--367, 2021.

\bibitem{20-jailbreak-llm}
X.~Shen, Z.~Chen, M.~Backes, Y.~Shen, and Y.~Zhang, ``"do anything now": Characterizing and evaluating in-the-wild jailbreak prompts on large language models,'' \emph{arXiv preprint arXiv:2308.03825}, 2023.

\bibitem{24-ed-guard-unsafe-llm}
S.~Goyal, M.~Hira, S.~Mishra, S.~Goyal, A.~Goel, N.~Dadu, D.~Kirushikesh, S.~Mehta, and N.~Madaan, ``{LLMGuard:} guarding against unsafe {LLM} behavior,'' in \emph{Proceedings of the AAAI Conference on Artificial Intelligence}, vol. 38(21), 2024, pp. 23\,790--23\,792.

\bibitem{6-Code-help-llm-guardrails}
M.~Liffiton, B.~E. Sheese, J.~Savelka, and P.~Denny, ``Codehelp: Using large language models with guardrails for scalable support in programming classes,'' in \emph{Proceedings of the 23rd Koli Calling International Conference on Computing Education Research}, Finland, 2024, pp. 1--11.

\bibitem{30-ed-socio-technical-safety-GenAI}
L.~Weidinger, M.~Rauh, N.~Marchal, A.~Manzini, L.~A. Hendricks, J.~Mateos-Garcia, S.~Bergman, J.~Kay, C.~Griffin, B.~Bariach \emph{et~al.}, ``Sociotechnical safety evaluation of generative {AI} systems,'' \emph{arXiv preprint arXiv:2310.11986}, 2023.

\bibitem{formalizing-fairness-in-ML}
P.~Gajane and M.~Pechenizkiy, ``On formalizing fairness in prediction with machine learning,'' \emph{arXiv preprint arXiv:1710.03184}, 2017.

\bibitem{generalizability-ai-model-2023}
W.~Li \emph{et~al.}, ``Segment anything model can not segment anything: Assessing {AI} foundation model’s generalizability in permafrost mapping,'' \emph{Remote Sensing}, vol.~16, no.~5, p. 797, 2024.

\bibitem{Customizing-fm-2024}
H.-I. Kim, K.~Yun, J.-S. Yun, and Y.~Bae, ``Customizing segmentation foundation model via prompt learning for instance segmentation,'' \emph{arXiv preprint arXiv:2403.09199}, 2024.

\bibitem{10-talking-abt-llm}
M.~Shanahan, ``Talking about large language models,'' \emph{Commun. ACM}, vol.~67, no.~2, p. 68–79, jan 2024.

\bibitem{22-ed-persuasion-challenge-llm}
Y.~Zeng, H.~Lin, J.~Zhang, D.~Yang, R.~Jia, and W.~Shi, ``How johnny can persuade {LLM}s to jailbreak them: Rethinking persuasion to challenge {AI} safety by humanizing {LLM}s,'' \emph{arXiv preprint arXiv:2401.06373}, 2024.

\bibitem{21-ed-llm-safety-against-prompt}
A.~Kumar, C.~Agarwal, S.~Srinivas, S.~Feizi, and H.~Lakkaraju, ``Certifying {LLM} safety against adversarial prompting,'' \emph{arXiv preprint arXiv:2309.02705}, 2023.

\bibitem{valmeekam2024llms}
K.~Valmeekam, K.~Stechly, and S.~Kambhampati, ``Llms still can't plan; can lrms? a preliminary evaluation of openai's o1 on planbench,'' \emph{arXiv preprint arXiv:2409.13373}, 2024.

\bibitem{Guaranteed-Safe-AI}
D.~Dalrymple \emph{et~al.}, ``Towards guaranteed safe {AI}: A framework for ensuring robust and reliable {AI} systems,'' \emph{arXiv preprint arXiv:2405.06624}, 2024.

\bibitem{1-ED-Guardrails-auto-ambigous-statement}
M.~Pawagi and V.~Kumar, ``Guardrails: Automated suggestions for clarifying ambiguous purpose statements,'' in \emph{Proceedings of the 16th Annual ACM India Compute Conference}, 2023, p. 55–60.

\bibitem{4-attack-to-llm}
N.~Mangaokar, A.~Hooda, J.~Choi, S.~Chandrashekaran, K.~Fawaz, S.~Jha, and A.~Prakash, ``{PRP}: Propagating universal perturbations to attack large language model guard-rails,'' \emph{arXiv preprint arXiv:2402.15911}, 2024.

\bibitem{lu2015tail}
Q.~Lu, X.~Xu, L.~Bass, L.~Zhu, and W.~Zhang, ``A tail-tolerant cloud api wrapper,'' \emph{IEEE Software}, vol.~32, no.~1, pp. 76--82, 2015.

\bibitem{RAG-attack-2024}
W.~Zou, R.~Geng, B.~Wang, and J.~Jia, ``{PoisonedRAG}: Knowledge poisoning attacks to retrieval-augmented generation of large language models,'' \emph{arXiv preprint arXiv:2402.07867}, 2024.

\bibitem{2-ed-toxicity-palm2}
A.~Khorramrouz, S.~Dutta, A.~Dutta, and A.~R. Khuda~Bukhsh, ``Down the toxicity rabbit hole: Investigating {PaLM 2} guardrails,'' \emph{arXiv preprint arXiv:2309.06415}, 2023.

\bibitem{Empirically-grounded-reference-architectures}
\BIBentryALTinterwordspacing
M.~Galster and P.~Avgeriou, ``Empirically-grounded reference architectures: a proposal,'' in \emph{Proceedings of the Joint ACM SIGSOFT Conference -- QoSA and ACM SIGSOFT Symposium -- ISARCS on Quality of Software Architectures -- QoSA and Architecting Critical Systems -- ISARCS}, ser. QoSA-ISARCS '11.\hskip 1em plus 0.5em minus 0.4em\relax New York, NY, USA: Association for Computing Machinery, 2011, p. 153–158. [Online]. Available: \url{https://doi.org/10.1145/2000259.2000285}
\BIBentrySTDinterwordspacing

\bibitem{wei2022chain}
J.~Wei, X.~Wang, D.~Schuurmans, M.~Bosma, F.~Xia, E.~Chi, Q.~V. Le, D.~Zhou \emph{et~al.}, ``Chain-of-thought prompting elicits reasoning in large language models,'' \emph{Advances in neural information processing systems}, vol.~35, pp. 24\,824--24\,837, 2022.

\bibitem{liu2023prompt}
Y.~Liu, G.~Deng, Y.~Li, K.~Wang, Z.~Wang, X.~Wang, T.~Zhang, Y.~Liu, H.~Wang, Y.~Zheng \emph{et~al.}, ``Prompt injection attack against llm-integrated applications,'' \emph{arXiv preprint arXiv:2306.05499}, 2023.

\bibitem{yohsua2024international}
\BIBentryALTinterwordspacing
B.~Yohsua, P.~Daniel, B.~Tamay, B.~Rishi, C.~Stephen, C.~Yejin, G.~Danielle, H.~Hoda, K.~Leila, L.~Shayne, M.~Vasilios, M.~Mantas, N.~Kwan~Yee, O.~Chinasa~T., R.~Deborah, S.~Theodora, T.~Florian, and M.~Soren, ``{International Scientific Report on the Safety of Advanced AI},'' {Department for Science, Innovation and Technology}, Tech. Rep., May 2024. [Online]. Available: \url{https://hal.science/hal-04612963}
\BIBentrySTDinterwordspacing

\bibitem{chen2024agentpoison}
Z.~Chen, Z.~Xiang, C.~Xiao, D.~Song, and B.~Li, ``Agentpoison: Red-teaming llm agents via poisoning memory or knowledge bases,'' \emph{arXiv preprint arXiv:2407.12784}, 2024.

\bibitem{sun2023aspect}
J.~Sun, Z.~Xing, X.~Xia, Q.~Lu, X.~Xu, and L.~Zhu, ``Aspect-level information discrepancies across heterogeneous vulnerability reports: Severity, types and detection methods,'' \emph{ACM Transactions on Software Engineering and Methodology}, vol.~33, no.~2, pp. 1--38, 2023.

\bibitem{dong2024taxonomy}
L.~Dong, Q.~Lu, and L.~Zhu, ``A taxonomy of agentops for enabling observability of foundation model based agents,'' \emph{arXiv preprint arXiv:2411.05285}, 2024.

\end{thebibliography}
\small



\end{document}